\documentclass[useAMS,usenatbib]{mn2e}

\usepackage{amsmath}
\usepackage{amssymb}
\usepackage{graphicx}
\usepackage{dcolumn}
\usepackage{bm}
\usepackage{ulem}

\title[Shear modulus of neutron star crust]{Shear modulus of neutron star crust}
\author[D. A. Baiko]{D. A. Baiko\thanks{E-mail:baiko@astro.ioffe.ru} \\
A. F. Ioffe Physical-Technical Institute, 
Politekhnicheskaya 26, 194021 St.-Petersburg, Russian Federation \\
Saint-Petersburg State Polytechnical University,
Politekhnicheskaya 29, 195251 St.-Petersburg, Russian Federation}

\begin{document}

\date{Accepted; Received ; in original form}

\pagerange{\pageref{firstpage}--\pageref{lastpage}} \pubyear{2011}

\maketitle

\label{firstpage}

\begin{abstract}
Shear modulus of solid neutron star crust is calculated by 
thermodynamic perturbation theory taking into account ion motion.  
At given density the crust is modelled as a body-centered cubic 
Coulomb crystal of fully ionized atomic nuclei of one type with 
the uniform charge-compensating electron background. 
Classic and quantum regimes of ion motion are considered. 
The calculations in the classic temperature range agree well with previous 
Monte Carlo simulations. At these temperatures the shear modulus 
is given by the sum of a positive contribution due to the static lattice 
and a negative $\propto T$ contribution due to the ion motion. The 
quantum calculations are performed for the first time. The main 
result is that at low temperatures the contribution to the shear modulus 
due to the ion motion saturates at a constant value, associated with zero-point 
ion vibrations. Such behavior is qualitatively similar to the zero-point 
ion motion contribution to the crystal energy.
The quantum effects may be important for lighter elements at higher densities,
where the ion plasma temperature is not entirely negligible compared
to the typical Coulomb ion interaction energy.
The results of numerical calculations are approximated by 
convenient fitting formulae. They should be used for precise
neutron star oscillation modelling, a rapidly developing branch 
of stellar seismology.
\end{abstract}


\section{Introduction}
\label{introduct}
Recent discovery of quasi-periodic oscillations (QPO) in soft gamma-repeaters 
\citep{IBS05,SW05,WS06} may be opening up an exciting possibility
into studying neutron stars by methods of seismology. The QPO are thought to be
related to neutron star vibrations and, in particular, originally, they were
thought to be related to torsional vibrations of neutron star crust
\citep{D98,P05}.

Even though it is now understood that the mechanism of neutron star oscillations 
is likely more complex and involves global oscillations of crust and core, coupled
by the frozen-in magnetic field \citep{L06,L07,L08}, it is still possible that the actual 
oscillation frequencies are related to pure crustal frequencies, the important
controlling factor being the magnetic field strength and geometry 
\citep[][and references therein]{GSA06,SW07,VHL10}. 
The crustal torsional vibration frequencies are determined by
the shear modulus of the solid neutron star crust. 
The main purpose of the present paper is to calculate this quantity.

The bulk of the neutron star crust is made of fully ionized ions (of varying
charge $Ze$ and mass $M$) in crystalline state, immersed in a nearly uniform strongly degenerate
electron gas. More specifically, the ions form a crystal, if the local temperature $T$ falls below 
the melting temperature $T_{\rm m} = Z^2 e^2 /(a \Gamma_{\rm m})$, where $\Gamma_{\rm m} \approx 175$,
and $a=(4 \pi n/3)^{-1/3}$ is the ion sphere radius ($n$ is the ion number density, $k_{\rm B}=1$).
Typically one assumes that the ion crystal is of body-centered cubic (bcc) type, as this structure
is preferable thermodynamically for strictly uniform electron background.

The state of the electron subsystem depends on matter density.  
We shall limit ourselves to such (not too low) densities, where electrons
are degenerate and ions are completely pressure ionized
\citep[$\rho \ga 10AZ$ g cm$^{-3}$, where $A$ is the ion mass number; 
see for discussion][]{PR95,HPY07}.
At these densities the model of uniform charge-compensating
background of electrons is reasonably good. It gets progressively
better with the growth of density 
becoming especially accurate
at $\rho \gg 10^6$ g cm$^{-3}$, where electrons are 
ultrarelativistic. The crystal of fully ionized ions with the uniform background 
of electrons is known as the Coulomb crystal.

In inner crust, at densities above the neutron drip density 
$\rho_{\rm d} \approx 4.3 \cdot 10^{11}$ g cm$^{-3}$, in addition to the Coulomb crystal 
of ions and electrons, there are neutrons
not bound in the atomic nuclei. The details of neutron interactions with nuclei 
are not known very well. 
It seems plausible that the properties of a strictly 
static crystal, i.e.\ a crystal with nuclei fixed at their lattice nodes, are 
determined by Coulomb forces. By contrast, the motion of nuclei about the
lattice nodes may be affected by the presence of neutrons. Lacking a good model of 
this effect, we shall assume that it can be described by an effective nucleus mass and 
renormalized ion plasma frequency within the framework of the Coulomb
crystal model.

The main purpose of this paper is thus to study the shear modulus of the 
Coulomb crystal.
The groundwork for this problem was laid down by \citet{F36}, who calculated the
shear modulus of the static bcc Coulomb lattice. 
More recently, \citet{OI90} 
calculated the shear modulus of the bcc Coulomb crystal taking into account the
motion of ions about their lattice nodes. In that work shear modulus was found numerically
with the aid of Monte Carlo simulations \citep[{e.g.,}][]{BST66}. By the nature of the method,
the motion of ions was treated classically. \citet{SVHO91}
further remarked that quantum effects were not important due to the smallness of the ion
plasma temperature compared to the typical lattice electrostatic energy. 
Finally, \citet{HH08} calculated the shear modulus of the Coulomb
crystal taking into account weak electron screening in the Thomas-Fermi model.
This calculation was done numerically using molecular dynamics method.
Again the motion of ions was strictly classic.

In this paper we shall use the thermodynamic perturbation theory 
to find the shear 
modulus of the Coulomb crystal with ion motion included
in the harmonic lattice model framework. 
Unlike numerical methods of \citet{OI90} and \citet{HH08} 
this approach is capable of tackling quantum effects. The quantum effects
are known to be important especially for lighter elements at higher densities. 
Quantum effects in the problem of Coulomb crystal elastic moduli are 
studied for the first time.

In addition to the outermost envelope of the external crust, the Coulomb crystal model fails
in the `nuclear pasta' region of the inner crust at densities
$\rho \ga 10^{14}$ g cm$^{-3}$, where nuclei become nonspherical.
Estimates of the shear modulus in this layer were reported by \citet{PP98}. 

Besides neutron star crusts, the Coulomb crystals are expected to form 
in solid cores of white dwarfs, to which the present results thus also apply.

\section{General theory}
\label{general}
A Coulomb crystal is composed of ions with charge $Ze$, arranged in a crystal 
lattice with equilibrium lattice sites $\bm{R}_I$, immersed in a rigid background
of electrons (charge $-e$). Background volume element coordinates are denoted as $\bm{r}$.
Suppose the crystal is deformed uniformly. Then the lattice remains perfect, but its 
equilibrium nodes move to new locations $\bm{X}_I$. The background volume 
element $\bm{r}$ shifts to a new position $\bm{x}$.  Using the 
displacement gradients
$u_{\alpha \beta} \equiv \partial X_\alpha / \partial R_\beta - \delta_{\alpha \beta}$
(for uniform deformations $u_{\alpha \beta} = {\rm const}$), 
one can express the coordinates of the new positions
via those of the old ones as
\begin{eqnarray}
     X_\alpha &=& R_\alpha + u_{\alpha \beta} R_\beta~, ~~~
\label{X} \\
     x_\alpha &=& r_\alpha + u_{\alpha \beta} r_\beta~.
\label{x} 
\end{eqnarray}
In this case, greek indices denote Cartesian coordinates. 
We do not distinguish between
upper and lower indices, and summation over repeated 
greek indices from 1 to 3 is always assumed.
The potential energy $U$ of a uniformly deformed Coulomb crystal 
can be written as
\begin{eqnarray}
     \frac{U}{Z^2 e^2} &=& \frac{1}{2} 
  \sideset{}{'}\sum_{IJ} \frac{1}{|\bm{X}_I + 
       \bm{u}_I - \bm{X}_J - \bm{u}_J|} 
\nonumber \\
    &-& \sum_{I} 
         \int  \frac{n~{\rm d} \bm{r}}{|\bm{X}_I+\bm{u}_I - \bm{x}|}
       + \frac{n^2}{2} \iint  
       \frac{{\rm d}\bm{r}_1 {\rm d}\bm{r}_2}{|\bm{x}_1 - \bm{x}_2|}~,
\label{Uini} 
\end{eqnarray}
where $\bm{u}_I$ is the $I$-th ion deviation from its deformed 
equilibrium position $\bm{X}_I$ 
due to thermal and zero-point vibrations, and $n$ is the 
ion number density in the nondeformed configuration $\{\bm{R}\}$. 
Integrations are over the nondeformed crystal volume $V$, prime means
that terms with $I=J$ are omitted.
Since a crystal lattice
realizes a local energy minimum with respect to small ion deviations from
the lattice nodes,
the energy $U$ can be approximately 
expressed as
\begin{eqnarray}
       U &\approx& U^{\rm st}(\{\bm{X}\}) + \delta U~,
\label{U0dU}
\end{eqnarray}
where $U^{\rm st}(\{\bm{X}\})$ is the potential energy of the static deformed lattice 
(i.e.\ the energy of the 
lattice with all ions located at the lattice nodes),
and $\delta U$ is the second order term of the Taylor expansion in powers 
of $\bm{u}_I$:
\begin{eqnarray}
    \delta U &=& \frac{1}{2} 
       \sum_{IJ} 
           \frac{\partial^2 U}{\partial u_I^\mu \partial u_J^\nu} 
          u_I^\mu u_J^\nu  \equiv \frac{1}{2} \sum_{IJ} 
           U_{IJ}^{\mu \nu} u_I^\mu u_J^\nu~,
\label{dltU} \\
           \frac{U_{IJ}^{\mu \nu}}{Z^2e^2} &=&
         \frac{\partial^2}{\partial X_I^\mu \partial X_I^\nu} \left\{
        \frac{\delta_{IJ}-1}{|\bm{X}_I - \bm{X}_J|} +  \sum_{K \ne I} 
         \frac{\delta_{IJ}}{|\bm{X}_I - \bm{X}_K|} \right.
\nonumber \\
    &-&   \left. \delta_{IJ} \int 
        \frac{n \, {\rm d}\bm{r} }{|\bm{X}_I - \bm{x}|} \right\}~.
\label{UKLab}
\end{eqnarray}
The last term in the curly brackets serves to cancel an infinity arising 
in the second one.

The potential energy of the static deformed lattice can be expanded in powers of the
displacement gradients as follows
\begin{equation}
           \frac{1}{V} U^{\rm st}(\{\bm{X}\}) = \frac{1}{V} U^{\rm st}(\{\bm{R}\}) + S^{\rm st}_{\alpha \beta} 
           u_{\alpha \beta} + 
           \frac{1}{2} S^{\rm st}_{\alpha \beta \gamma \lambda} u_{\alpha \beta} 
          u_{\gamma \lambda}.
\label{U0expansion}
\end{equation}
In this case, $U^{\rm st}(\{\bm{R}\})$ is the potential energy of the static 
nondeformed lattice, while the first order expansion coefficient
$S^{\rm st}_{\alpha \beta}$ can be expressed via electrostatic crystal
pressure: $S^{\rm st}_{\alpha \beta} = - P^{\rm st} \delta_{\alpha \beta}$
[note, that to first order 
$\delta V/V = u_{\alpha \alpha}$ for an arbitrary deformation, where
$\delta V$ is the volume change due to the deformation;
also see Sect.\ \ref{pressure} and \citet{W67}]. 
$S^{\rm st}_{\alpha \beta \gamma \lambda}$ are the static 
lattice elastic 
coefficients. Elastic coefficients as second derivatives with 
respect to the displacement gradients were introduced 
by \citet{H50}. 

Since the static lattice coefficients are well known
\citep[][see also Sect.\ \ref{results}]{F36},
the main subject of the present paper is the
temperature and density dependences 
of the elastic coefficients associated with ion vibrations about
their lattice sites. Thus we shall focus on $\delta U$ 
of Eqs.\ (\ref{U0dU}) and (\ref{dltU}). 
Making use of the standard in solid-state theory
\citep[e.g.,][]{BH54} 
collective coordinates $A_{\bm{k}}^\mu$
(for brevity of notation we consider simple lattices only,
i.e.\ lattices with only one ion in the elementary cell):
\begin{eqnarray}
          u_I^\mu &=& \frac{1}{\sqrt{MN}} \sum_{\bm{k}} 
          A_{\bm{k}}^\mu \exp{({\rm i} \bm{k} \cdot \bm{R}_I)}~,
\label{uia} 
\end{eqnarray}
where wavevector $\bm{k}$ belongs to the first Brillouin zone 
of the nondeformed 
lattice, while $M$ and $N$ are the ion mass and total number,
$\delta U$ can be written as
\begin{eqnarray}
    \delta U &=& \frac{1}{2 MN} 
            \sum_{\bm{k} \bm{k}'}   A_{\bm{k}}^\mu
           A_{\bm{k}'}^\nu 
\nonumber \\
          &\times& \sum_{IJ} U_{IJ}^{\mu \nu} 
                 \exp{[{\rm i} \bm{k} \cdot (\bm{R}_I - \bm{R}_J) + 
                  {\rm i} (\bm{k} + \bm{k}') \cdot \bm{R}_J]}~.
\label{dltUnorm}
\end{eqnarray}
At fixed $J$, the sum over $I$ in Eq.\ (\ref{dltUnorm}), 
along with the sum over $K$ and the integral in Eq.\ (\ref{UKLab}),
can be extended to infinity. The remaining finite sum over $J$ 
in Eq.\ (\ref{dltUnorm})
produces $N \delta_{-\bm{k}\bm{k}'}$. 
%
Then $\delta U$ can be rewritten as
\begin{eqnarray}
    \delta U &=& \frac{Z^2e^2}{2M} \sum_{\bm{k}} A_{\bm{k}}^\mu  A_{-\bm{k}}^\nu
              \sum_{I \ne 0} 
                [1 - \exp{({\rm i} \bm{k} \cdot \bm{R}_I)}]
                \frac{\partial^2 X^{-1}_I}{\partial X^\mu_I \partial X^\nu_I}
\nonumber \\
             &\equiv& \frac{1}{2} \sum_{\bm{k}} 
           A_{\bm{k}}^\mu  A_{-\bm{k}}^\nu 
             D^{\mu \nu} (\bm{k}, \{\bm{X}\})~,
\label{dltUwork}
\end{eqnarray}
where the deformed dynamic matrix $D^{\mu \nu} (\bm{k}, \{\bm{X}\})$
has been introduced. That matrix can also be expanded
in powers of the displacement gradients and thus related to
higher order nondeformed lattice energy derivatives:
\begin{eqnarray}
             D^{\mu \nu} (\bm{k}, \{\bm{X}\}) &\approx& 
          D^{\mu \nu} (\bm{k}, \{\bm{R}\}) + 
             D_{\alpha \beta}^{\mu \nu} (\bm{k}) u_{\alpha \beta} 
\nonumber \\
            &+& 
           \frac{1}{2} D_{\alpha \beta \gamma \lambda}^{\mu \nu} 
           (\bm{k}) u_{\alpha \beta} u_{\gamma \lambda}~,
\label{Dexpansion} \\
             D_{\alpha \beta}^{\mu \nu} (\bm{k}) &=&
            \frac{Z^2e^2}{M}  
            \sideset{}{'}\sum_{\bm{R}} 
                   [1 - \exp{({\rm i} \bm{k} \cdot \bm{R})}] R_\beta 
\nonumber \\
            &\times&      \frac{\partial^3 R^{-1}}
                  {\partial R_\alpha \partial R_\mu \partial R_\nu}~,
\label{D(ab)} \\
            D_{\alpha \beta \gamma \lambda}^{\mu \nu} (\bm{k}) &=&  \frac{Z^2e^2}{M}
              \sideset{}{'}\sum_{\bm{R}} 
                   [1 - \exp{({\rm i} \bm{k} \cdot \bm{R})}] R_\beta R_\lambda 
\nonumber \\
          &\times& \frac{\partial^4 R^{-1}}
                  {\partial R_\alpha \partial R_\gamma \partial R_\mu \partial R_\nu}~.
\label{D(ab)(gl)}
\end{eqnarray}
The prime means that summations over ${\bm R}$ run over all nondeformed 
lattice vectors except $\bm{R}=0$. 
Practical expressions for the coefficients of the expansion
(\ref{Dexpansion}) can be found in Appendix A. They were obtained using
the standard Ewald technique \citep[e.g.,][]{BH54}. 
In Eqs.\ (\ref{dltUwork}), (\ref{D(ab)}), and 
(\ref{D(ab)(gl)}) it is understood that the apparent divergences are canceled
by the electron background term of Eq.\ (\ref{UKLab}). In practical formulae 
of Appendix A this is reflected by the absence of the ${\bm q}=0$ term in sums
over reciprocal lattice vectors.

Nondeformed dynamic matrix $D^{\mu \nu} (\bm{k}, \{\bm{R}\})$ is given by  
the expression for $D^{\mu \nu} (\bm{k}, \{\bm{X}\})$, stemming 
from Eq.\ (\ref{dltUwork}), with $X_I$ replaced by $R_I$.
This matrix determines
frequencies $\omega_{\bm{k}j}$ and polarization vectors $\bm{e}_{\bm{k}j}$
of nondeformed crystal oscillations: 
$D^{\mu \nu} (\bm{k}, \{\bm{R}\}) e^\nu_{\bm{k}j} = \omega^2_{\bm{k}j} e^\mu_{\bm{k}j}$,
where $j$ enumerates oscillation modes with given $\bm{k}$ 
($j=1,2,3$ for simple lattices). 

Let us expand $\bm{A}_{\bm{k}}$ over the basis of 
$\bm{e}_{\bm{k}j}$:   $A^\mu_{\bm{k}} = \sum_{j=1}^3 e^\mu_{\bm{k}j} Q_{\bm{k}j}$.
Then the oscillatory potential energy (\ref{dltUwork}) shall consist of three parts
$\delta U = H_0 + H_1 + H_2$, where
\begin{eqnarray}
           H_0 &=& \frac{1}{2} \sum_{\bm{k}j} \omega^2_{\bm{k}j}
           Q_{\bm{k}j} Q_{-\bm{k}j} 
\label{H0} \\  
           H_1 &=&   u_{\alpha \beta} 
           \sum_{\bm{k}jj'}  \frac{1}{2} 
           \mathit{\Phi}^{jj'}_{\alpha \beta} (\bm{k})
           Q_{\bm{k}j} Q_{-\bm{k}j'}
\label{H1} \\
           H_2 &=& \frac{1}{2} \, 
         u_{\alpha \beta} u_{\gamma \lambda}
           \sum_{\bm{k}jj'}  \frac{1}{2} 
           \mathit{\Phi}^{jj'}_{\alpha \beta \gamma \lambda} (\bm{k})
           Q_{\bm{k}j} Q_{-\bm{k}j'}~, 
\label{H2}
\end{eqnarray}
and, following \citet{BH54},
we have introduced quantities
\begin{eqnarray}
                  \mathit{\Phi}^{jj'}_{\alpha \beta} (\bm{k}) &=&  
                 e^\mu_{\bm{k}j} e^\nu_{-\bm{k}j'} 
                  D_{\alpha \beta}^{\mu \nu} (\bm{k})~,   
\label{Phi(ab)} \\
                  \mathit{\Phi}^{jj'}_{\alpha \beta \gamma \lambda} 
               (\bm{k}) &=&  
                 e^\mu_{\bm{k}j} e^\nu_{-\bm{k}j'} 
                  D_{\alpha \beta \gamma \lambda}^{\mu \nu} (\bm{k})~.   
\label{Phi(ab)(gl)}
\end{eqnarray}
In this case, $H_0$ is the oscillatory potential energy of the nondeformed
lattice, while $H_1$ and $H_2$ represent a perturbation of this quantity
due to the deformation. 

In quantum mechanics coordinates $Q_{\bm{k}j}$ become operators. It is convenient to
switch to second quantization representation, where operators $Q_{\bm{k}j}$ are expressed
via phonon creation and annihilation operators $a_{\bm{k}j}^\dagger$ and $a_{\bm{k}j}$:
\begin{eqnarray}
          Q_{\bm{k}j} &=& \sqrt{\frac{\hbar}{2 \omega_{\bm{k}j}}} 
          (a_{\bm{k}j} + a_{-\bm{k}j}^\dagger)~,
\nonumber \\
          Q_{-\bm{k}j} &=& \sqrt{\frac{\hbar}{2 \omega_{\bm{k}j}}} 
            (a_{-\bm{k}j} + a_{\bm{k}j}^\dagger)~.
\label{Qva+a}
\end{eqnarray}
It is now possible to obtain expansion of the free energy in powers of the displacement gradients
using the thermodynamic perturbation theory \citep[e.g.,][]{LL80}:
\begin{eqnarray}
    \delta F &=& \sum_{n} V_{nn} w_n 
    + \sideset{}{'}\sum_{n,m} \frac{|V_{nm}|^2 w_n}
     {E_n^{(0)} - E_m^{(0)}} 
\nonumber \\
    &+& \frac{1}{2T} \left[ \left( \sum_{n} V_{nn} w_n \right)^2
      - \sum_{n} (V_{nn})^2 w_n \right] + \ldots 
\label{deltaF}
\end{eqnarray}
In this case, $V=H_1+H_2$ is the perturbation operator, indices $n$ and $m$ run 
over all possible unperturbed quantum states (which in second quantization means 
a sum over all possible
phonon occupation numbers in all modes), and $w_n = \exp{\{(F_0 - E_n^{(0)})/T\}}$
is the probability of the quantum state $n$, $F_0$ and $E_n^{(0)}$ being unperturbed
free energy and quantum state energy. Terms with $n=m$ in the second sum are excluded.

For simple lattices, $\bm{e}_{-\bm{k}j} = \bm{e}_{\bm{k}j}$, 
while all matrices $D$ on the r.\ h.\ s.\ of Eq.\ (\ref{Dexpansion}), 
are real and symmetric with respect to their upper indices $\mu$ and $\nu$.
Consequently, Eq.\ (\ref{deltaF}) can be written as  
$(\delta F)/V = S^{\rm ph}_{\alpha \beta} u_{\alpha \beta} + 
0.5 S^{\rm ph}_{\alpha \beta \gamma\lambda} u_{\alpha \beta} u_{\gamma \lambda} + \ldots ~$,
with 
\begin{eqnarray}
   S^{\rm ph}_{\alpha \beta} &=&  \frac{1}{2V} \sum_{\bm{k}j} 
           \mathit{\Phi}^{jj}_{\alpha \beta} 
           \frac{\hbar}{2 \omega_{\bm{k}j}} (1+ 2 \bar{n}_{\bm{k}j})~,
\label{STab} \\
   S^{\rm ph}_{\alpha \beta \gamma \lambda} &=&  \frac{1}{2V} \sum_{\bm{k}j}
            \frac{\hbar}{2 \omega_{\bm{k}j}} (1+ 2 \bar{n}_{\bm{k}j})
\nonumber \\ &\times&
            \left[ \mathit{\Phi}^{jj}_{\alpha \beta \gamma \lambda} -
            \frac{\mathit{\Phi}^{jj}_{\alpha \beta} 
       \mathit{\Phi}^{jj}_{\gamma \lambda}}{2 \omega^2_{\bm{k}j}}
     + 2 \sum_{j' \ne j}
                \frac{\mathit{\Phi}^{jj'}_{\alpha \beta} 
          \mathit{\Phi}^{jj'}_{\gamma \lambda}}{\omega^2_{\bm{k}j} - \omega^2_{\bm{k}j'}}
             \right]  
\nonumber \\
     &-& \frac{1}{2V} \sum_{\bm{k}j} \frac{\hbar^2}{T} 
             (\bar{n}_{\bm{k}j}+1) \bar{n}_{\bm{k}j} \frac{\mathit{\Phi}^{jj}_{\alpha \beta} 
          \mathit{\Phi}^{jj}_{\gamma \lambda}}{2 \omega^2_{\bm{k}j}}~,
\label{STabgl}
\end{eqnarray}
where  $\bar{n}_{\bm{k}j} = [\exp{(\hbar \omega_{\bm{k}j}/T)}-1]^{-1}$
is the average occupation number in a Bose system, and the argument $\bm{k}$ 
is implicit for all $\mathit{\Phi}$'s. 
Note a typo in eq.\ (41.38) of \citet{BH54}, which
differs from our expression (\ref{STabgl}) 
by the absence of a 
factor 2 in front of the $\sum_{j'\ne j}$ in square brackets.
The term with the double sum over $j$ and $j'$ in Eq.\ (\ref{STabgl})
can be also written as
\begin{equation}
  \frac{1}{V} \sum_{\bm{k}j, j'>j} \mathit{\Phi}^{jj'}_{\alpha \beta} 
          \mathit{\Phi}^{jj'}_{\gamma \lambda} 
           \frac{q^2_{\bm{k}j} - q^2_{\bm{k}j'}}{\omega^2_{\bm{k}j} - \omega^2_{\bm{k}j'}}
           ~~~ {\rm with} ~~~ 
          q^2_{\bm{k}j} \equiv \frac{\hbar}{2 \omega_{\bm{k}j}} (1+ 2 \bar{n}_{\bm{k}j})~,
\nonumber
\end{equation}
which removes the singularity associated with
degenerate phonon modes $\omega_{\bm{k}j} = \omega_{\bm{k}j'}$.

The expansion of $\delta F$ in powers of displacement gradients
is done at fixed temperature. Accordingly, the elastic 
coefficients of Eq.\ (\ref{STabgl}) are isothermal. Adiabatic elastic coefficients 
are defined via expansion of energy in powers of displacement gradients at fixed 
entropy (see Sect.\ \ref{SnT}).

Besides the Huang expansion in powers of the displacement gradients it is customary to 
expand thermodynamic potentials in powers of the Lagrangian strain parameters
\begin{eqnarray}
               \eta_{\alpha \beta} &=& \frac{1}{2} 
              (u_{\alpha \beta} + u_{\beta \alpha}
               +  u_{\lambda \alpha} u_{\lambda \beta})~.
\label{bar_u}
\end{eqnarray}
In this way one obtains $(\delta F)/V = C_{\alpha \beta} \eta_{\alpha \beta} + 
0.5 C_{\alpha \beta \gamma\lambda} \eta_{\alpha \beta} \eta_{\gamma \lambda} + \ldots~$
Since expansions in terms of $u_{\alpha \beta}$ and $\eta_{\alpha \beta}$ must coincide, 
one has \citep{W67}
\begin{eqnarray}
   C_{\alpha \beta} &=& S_{\alpha \beta}~,
\nonumber \\
   C_{\alpha \beta \gamma \lambda} &=& S_{\alpha \beta \gamma \lambda}
          - \delta_{\alpha \gamma} S_{\beta \lambda}~,
\label{CTabgl}
\end{eqnarray}
with the same relationships holding for partial contributions (e.g., `st', `ph', etc).
Coefficients $C$ have complete Voigt symmetry, that is 
$C_{\alpha \beta \gamma \lambda} = C_{\beta \alpha \gamma \lambda} = C_{\gamma \lambda \alpha \beta}$.
In general, this is not the case for the $S$-coefficients. In cubic symmetry there are only
three nontrivial $C$-coefficients $C_{1111}$, $C_{1122}$, and $C_{1212}$. In Voigt notation they are known as 
$C_{11}$, $C_{12}$, and $C_{44}$, respectively.

The results of numerical calculations of elastic coefficients
are presented in Sect.\ \ref{results}.

\section{Relation to phonon pressure}
\label{pressure}
As shown, e.g., by \citet{W67}, in the presence of an initial stress in the nondeformed
configuration, $-S_{\alpha \beta}$ (whether isothermal or adiabatic)
is equal to this stress. In the case of a Coulomb crystal in neutron star crust such initial 
stress is produced by pressure. Consequently, 
$S_{\alpha \beta} = -P \delta_{\alpha \beta}$, and
$S_{\alpha \beta \gamma \lambda} = - P \delta_{\alpha \gamma} \delta_{\beta \lambda} +  C_{\alpha \beta \gamma \lambda}$. 
It is thus clear that while $C_{1212}=C_{1221}$, $S_{1212} \ne S_{1221} = S_{1212}+P$.

If the total free energy
is a sum of several partial contributions, their first derivatives yield
partial pressures. 
Just like $S^{\rm st}_{\alpha \beta}$ in Eq.\ (\ref{U0expansion}) is related to electrostatic 
crystal pressure, $S^{\rm ph}_{\alpha \beta} = -P^{\rm ph} \delta_{\alpha \beta}$, where $P^{\rm ph}$ 
is the phonon pressure. Phonon pressure 
is found as a volume derivative of the phonon thermodynamic potential $\Omega^{\rm ph}$:
\begin{eqnarray}
          P^{\rm ph} &=& - \left(\frac{\partial \Omega^{\rm ph}}{\partial V} \right)_{\mu,T}~,
\label{Pph} \\
         \Omega^{\rm ph} &=& \sum_{\bm{k}j} \left\{ \frac{\hbar \omega_{\bm{k}j}}{2} + 
        T \ln{\left[ 1-\exp{\left(- \frac{\hbar \omega_{\bm{k}j}}{T} \right)}\right]}
         \right\}~.
\label{Omph}
\end{eqnarray}
The first and second terms in Eq.\ (\ref{Omph}) describe zero-point and 
thermal motion, respectively. The volume dependence is contained only in phonon frequencies. 
The ratios of phonon frequencies to the ion plasma frequency 
$\omega_{\rm p} = \sqrt{4 \pi n Z^2 e^2 / M}$
are universal functions for a given lattice type, and thus 
$\omega_{\bm{k}j} \propto \omega_{\rm p} \propto n^{1/2} \propto V^{-1/2}$.
It follows, that 
\begin{eqnarray}
   P^{\rm ph} &=& \frac{1}{2V} \sum_{\bm{k}j} \left\{ \frac{\hbar \omega_{\bm{k}j}}{2} + 
        \frac{\hbar \omega_{\bm{k}j}}{\exp{(\hbar \omega_{\bm{k}j}/T)} - 1}
         \right\} 
\nonumber \\
    &=& \frac{1}{4V} \sum_{\bm{k}j} \hbar \omega_{\bm{k}j} 
          ( 1 + 2 \bar{n}_{\bm{k}j} )~.
\label{P-deriv}
\end{eqnarray}
We can, therefore, assert the following identity:
\begin{equation}
      \sum_{\bm{k}j} \left( \frac{\mathit{\Phi}^{jj}_{\alpha \beta}}{\omega_{\bm{k}j}} + 
      \omega_{\bm{k}j} \delta_{\alpha \beta} \right)
          ( 1 + 2 \bar{n}_{\bm{k}j} ) = 0~.
\label{identity}
\end{equation}
This identity can be proven directly (at least for bcc lattice). First, we notice that from 
Eq.\ (\ref{Phi(ab)}) together with explicit formulae (\ref{DR(ab)}) and (\ref{DG(ab)})
from Appendix A, it is clear that  
$\mathit{\Phi}^{jj}_{xy}(\bm{k}_1)= - \mathit{\Phi}^{jj}_{xy}(\bm{k}_2)$, where
$\bm{k}_1$ and $\bm{k}_2$ differ from each other by the sign of their $x$-coordinate 
(or $y$-coordinate) and likewise for $\mathit{\Phi}^{jj}_{\alpha \beta}$ with other pairs 
of indices $\alpha \ne \beta$. By contrast, 
$\mathit{\Phi}^{jj}_{xx}$, $\mathit{\Phi}^{jj}_{yy}$, and 
$\mathit{\Phi}^{jj}_{zz}$ do not change under a sign change of any of their argument coordinates. 
If, on the other hand, $\bm{k}_1$ and $\bm{k}_2$ differ by the interchange of $x$- and 
$y$-coordinates, then $\mathit{\Phi}^{jj}_{xx}(\bm{k}_1)= \mathit{\Phi}^{jj}_{yy}(\bm{k}_2)$,
$\mathit{\Phi}^{jj}_{zz}(\bm{k}_1)= \mathit{\Phi}^{jj}_{zz}(\bm{k}_2)$, and likewise for
interchange of $x$- and $z$- or $y$- and $z$-coordinates.
This means that 
$\sum_{i=1}^{48} \mathit{\Phi}^{jj}_{\alpha \beta}(\bm{k}_i) 
= 16  [\mathit{\Phi}^{jj}_{xx}(\bm{k}) + \mathit{\Phi}^{jj}_{yy}(\bm{k}) + 
\mathit{\Phi}^{jj}_{zz}(\bm{k})] \delta_{\alpha \beta}$. 
The sum on the l.\ h.\ s.\ is over 48 Brillouin zone vectors
(with identical length $|\bm{k}|$), obtained from
$\bm{k}$ by 6 possible permutations of absolute values of 
its Cartesian coordinates and, for each permutation, by 8 possible 
combinations of signs assigned to those coordinates.

For uniform compression $\delta \omega_{\bm{k}j} = - \omega_{\bm{k}j} \delta V /(2V)$ and
$u_{\alpha \beta} = \delta_{\alpha \beta} \, \delta V /(3V)$.
On the other hand, 
$\omega^2_{\bm{k}j} =  e^\mu_{\bm{k}j} e^\nu_{-\bm{k}j} D^{\mu \nu} (\bm{k},\{\bm{R}\})$,
and therefore, $\delta \omega^2_{\bm{k}j} = u_{\alpha \beta} 
\mathit{\Phi}^{jj}_{\alpha \beta}$, because
variation of a polarization vector must be orthogonal to it 
in order to maintain its unit length. Combining these results we obtain 
$\mathit{\Phi}^{jj}_{\gamma \gamma} = -3 \omega^2_{\bm{k}j}$,
which proves Eq.\ (\ref{identity}). It is obvious from the derivation, 
that in place of $( 1 + 2 \bar{n}_{\bm{k}j} )$ we can have 
an arbitrary function of $|\bm{k}|$ and $j$.

\section{Effective shear modulus}
\label{average}
Free energy expansion coefficients, introduced in Sect.\ \ref{general},
also determine elastic stress tensor of deformed crystal. If deformation with
displacement gradient $u_{\alpha \beta}$ is applied to a configuration under
initial isotropic pressure $P$, the stress tensor 
$\sigma_{\alpha \beta}$, equal initially to 
$- P \delta_{\alpha \beta}$, will change by \citep{W67}
\begin{eqnarray}
   \delta \sigma_{\alpha \beta} &=& \frac{1}{2} B_{\alpha \beta \gamma \lambda} 
       (u_{\gamma \lambda} + u_{\lambda \gamma})~,
\label{dsigma} 
\end{eqnarray}
where
\begin{eqnarray}
   B_{\alpha \beta \gamma \lambda} = S_{\alpha \beta \gamma \lambda} - 
  P (\delta_{\alpha \lambda} \delta_{\beta \gamma} - \delta_{\alpha \beta} \delta_{\gamma \lambda})~.
\label{BvsS}
\end{eqnarray}
Thus $B_{1111} = S_{1111}$, $B_{1122} = S_{1122}+P$, $B_{1212} = S_{1212} = B_{1221}$.

Expression (\ref{dsigma}) allows one to write down linearized elastic medium equation of motion. 
In nonuniform matter $\delta \sigma_{\alpha \beta}$
of Eq.\ (\ref{dsigma}) gives Lagrangian variation of the stress tensor.
In realistic neutron star modelling, equation of motion must also take into account magnetic field and
non-uniformity of matter and initial stress, associated with gravitation 
[cf., e.g., eq.\ (9) in \citet{Cetal86}]. If all 
these complications are omitted, the equation of motion reads [eq.\ (2.23) of \citet{W67}]
\begin{eqnarray}
  \rho \ddot{u}_\alpha = B_{\alpha \beta \gamma \lambda} 
  \frac{\partial^2 u_\gamma}{\partial R_\beta \partial R_\lambda}
 = S_{\alpha \beta \gamma \lambda} 
  \frac{\partial^2 u_\gamma}{\partial R_\beta \partial R_\lambda}~,
\label{eq_motion}
\end{eqnarray}
where $\rho$ is the nondeformed mass density and $\bm{u}$ is the displacement
(so that $\bm{X}=\bm{R}+\bm{u}$). 

In Fourier space one has
\begin{eqnarray}
      \rho \omega^2 u^2 &=& S_{\alpha \beta \gamma \lambda} u_\alpha u_\gamma k_\beta k_\lambda~,
\label{eq_motion_Fourier} 
\end{eqnarray}
which for cubic symmetry can be expanded as
\begin{eqnarray}
      \rho \omega^2 u^2 &=& S_{1111} (u_x^2 k_x^2 +u_y^2 k_y^2 +u_z^2 k_z^2)
\nonumber \\ 
          &+& 2 S_{1122}  (u_x k_x u_y k_y + u_x k_x u_z k_z + u_z k_z u_y k_y)
\nonumber \\
          &+& 2 S_{1221}  (u_x k_x u_y k_y + u_x k_x u_z k_z + u_z k_z u_y k_y)
\nonumber \\
          &+& S_{1212} (u_x^2 k_y^2+u_y^2 k_x^2+u_x^2 k_z^2
\nonumber \\
   &+& u_z^2 k_x^2+u_y^2 k_z^2+u_z^2 k_y^2)~.
\label{cub_expl}
\end{eqnarray}

From Eqs.\ (\ref{dsigma}) and (\ref{cub_expl}) it is clear that in a perfect crystal 
it is $B_{1212} = S_{1212}$ that produces a response to a shear deformation.
For this reason we shall call $S_{1212}$ the {\it elastic shear coefficient}. However,
it was proposed by \citet{OI90} that in order to describe transverse modes in neutron star crusts, presumably
composed of many small randomly oriented crystalline domains,
one has to consider a directional average of the above equation. In particular, one has to average over
directions of $\bm{u}$ perpendicular to $\bm{k}$ and then over all possible directions of $\bm{k}$.
The resulting isotropic phase velocity $\omega/k$ should then be 
equated to effective shear wave speed $\sqrt{\mu_{\rm eff}/\rho}$, 
where $\mu_{\rm eff}$ is the effective shear modulus. The Lagrangian stress tensor variation is then 
approximated by the isotropic medium expression [e.g., eq.\ (15) in \citet{Cetal86}] with
$\mu_{\rm eff}$ in place of the shear modulus.

The necessary averaging is easy to carry out. Firstly,
\begin{eqnarray}
    \langle u_\alpha u_\beta \rangle &=& \frac{u^2}{2} \left( \delta_{\alpha \beta}
           - \frac{k_\alpha k_\beta}{k^2} \right) ~.
\label{<u>}
\end{eqnarray} 
And secondly, averaging over angles of 
$\bm{k}$ yields
\begin{eqnarray}
           \frac{\rho \omega^2}{k^2} = \frac{1}{5}(S_{1111}-S_{1122}-S_{1221}+4 S_{1212})~.
\label{mueff}
\end{eqnarray}
The combination on the r.\ h.\ s.\ of Eq.\ (\ref{mueff}) is just the {\it effective shear modulus}
$\mu_{\rm eff}$ in question.

It is not quite clear whether this averaging oversimplifies the real situation in neutron star 
crust. Firstly, it is not known how small and randomly oriented the crystalline domains making up
the crust really are, and whether or not one should consider instead a more regular crystal structure. 
For instance, as shown by \citet{B09}, bcc Coulomb crystal in magnetic field has minimum energy,
if it is oriented so that the direction of the magnetic field coincides with the direction towards 
one of the nearest neighbors. This effect is due to a dependence of zero-point energy 
on mutual orientation of the magnetic field and crystal axes. So, it is easy to imagine that
during star cooling the crust solidifies in such a way that the 
direction towards a nearest neighbor coincides with that of the magnetic field. This will 
produce a large scale ordered crystal structure. Secondly, if we agree
with the notion of crust as a collection of small randomly oriented domains,
it is not Eq.\ (\ref{eq_motion}) that has to be averaged, but the full 
equation of motion, which differs from (\ref{eq_motion}) by the presence 
of important anisotropies due to magnetic field and gravitation.

Since $\mu_{\rm eff}$ contains all elastic coefficients
(e.g., $S_{1111}$ which is related to bulk compressibility) 
one may wonder whether
any other subsystem, besides the ion lattice, contributes to $\mu_{\rm eff}$.
If a partial contribution to the energy (or the free energy) 
is a function of particle density only (e.g.\ kinetic energy of the
degenerate electron gas), the Huang coefficients, associated with it, 
can be written as 
%
\begin{eqnarray}
    VS_{\alpha \beta \gamma \lambda} = 
       \frac{\partial^2 U}{\partial u_{\alpha \beta} 
        \partial u_{\gamma \lambda}}
       &=& \frac{\partial^2 n_{\rm x}}
        {\partial u_{\alpha \beta} \partial u_{\gamma \lambda}}
          \frac{\partial U}{\partial n_{\rm x}} 
\nonumber \\
      &+& 
          \frac{\partial n_{\rm x}}{\partial u_{\alpha \beta}}
          \frac{\partial n_{\rm x}}{\partial u_{\gamma \lambda}}
          \frac{\partial^2 U}{\partial n_{\rm x}^2}~.
\label{S_isotr}
\end{eqnarray}
In this case, $n_{\rm x}$ is the density in the deformed configuration
which, to second order in displacement gradients, reads
\begin{eqnarray}
  n_{\rm x} &=& \frac{n}{ {\rm det}(1+u_{\alpha \beta})} 
       \approx n \left[ 1 - {\rm Tr}(u_{\alpha \beta}) \right.
\nonumber \\
       &+& u_{11}^2 + u_{22}^2 + u_{33}^2 
       +  u_{11} u_{22} + u_{11} u_{33}  
\nonumber \\
       &+& \left.  u_{22} u_{33} 
       +  u_{13} u_{31}
       + u_{12} u_{21} + u_{23} u_{32} \right]~,
\label{n-deform}
\end{eqnarray}
where $n$ is the nondeformed density. 
Consequently, one finds
\begin{eqnarray}
     VS_{1111} &=& 2 n \frac{\partial U}{\partial n_{\rm x}} +
                   n^2 \frac{\partial^2 U}{\partial n_{\rm x}^2}~,
\nonumber \\
     VS_{1122} &=&  n \frac{\partial U}{\partial n_{\rm x}} +
                   n^2 \frac{\partial^2 U}{\partial n_{\rm x}^2}~,
\nonumber \\
     VS_{1221} &=& n \frac{\partial U}{\partial n_{\rm x}}~,
\nonumber \\
     VS_{1212} &=& 0~.
\label{S_isotr_spec}     
\end{eqnarray}
The important implication is that 
there are no partial contributions to neither 
$S_{1212}$ nor $\mu_{\rm eff}$ [see Eq.\ (\ref{mueff})] due to such 
partial contributions to the (free) energy. 
In particular, neither electrons nor dripped neutrons 
\citep[in the standard model of neutron gas, e.g.,][]{ST83} in neutron 
star crust contribute to the effective shear modulus.

\section{Isothermal and adiabatic elastic coefficients}
\label{SnT}
In the previous sections we have found formulae for isothermal Huang coefficients 
$S^{\rm ph}_{\alpha \beta \gamma\lambda}$ and effective shear modulus 
$\mu^{\rm ph}_{\rm eff} = (S^{\rm ph}_{1111}-S^{\rm ph}_{1122}-S^{\rm ph}_{1221}+4 S^{\rm ph}_{1212})/5$.
Adiabatic Huang coefficients may be defined in the same way, the only difference 
being that the energy is expanded in powers of displacement gradients $u_{\alpha \beta}$ 
(instead of the free energy).
Adiabatic coefficients are likely much more appropriate for neutron star seismology.
In this section we show that isothermal and adiabatic $S_{1212}$ as well as $\mu_{\rm eff}$ are 
actually the same. In order to prove this, we note that
\begin{eqnarray}
    \left(\frac{\partial E}{\partial u_{\alpha \beta}} \right)_S = 
       \frac{\partial (E,S)}{\partial (u_{\alpha \beta},T)} \, 
       \frac{\partial (u_{\alpha \beta},T)}{\partial (u_{\alpha \beta},S)}
       ~~~~~~~~~~~~~~~~~~~~~~~~~ &&
\nonumber \\
   = \left(\frac{\partial E}{\partial u_{\alpha \beta}} \right)_T - 
         \left(\frac{\partial E}{\partial T} \right)_{u_{\alpha \beta}}      
         \left(\frac{\partial S}{\partial u_{\alpha \beta}} \right)_T      
         \left(\frac{\partial S}{\partial T} \right)^{-1}_{u_{\alpha \beta}}  &&    
\nonumber \\
   =  \left(\frac{\partial E}{\partial u_{\alpha \beta}} \right)_T +     
         \left(\frac{\partial E}{\partial T} \right)_{u_{\alpha \beta}}      
         \left(\frac{\partial T}{\partial u_{\alpha \beta}} \right)_S~. ~~~~~~~~~~~~&&
\label{aux_derive}
\end{eqnarray}
Substituting
\begin{equation}
         \left(\frac{\partial E}{\partial T} \right)_{u_{\alpha \beta}}  =
         \left(\frac{\partial E}{\partial S} \right)_{u_{\alpha \beta}}      
         \left(\frac{\partial S}{\partial T} \right)_{u_{\alpha \beta}}
\end{equation}
into Eq.\ (\ref{aux_derive}) we see that
\begin{eqnarray}
         \left(\frac{\partial E}{\partial u_{\alpha \beta}} \right)_S +
         \left(\frac{\partial E}{\partial S} \right)_{u_{\alpha \beta}}
         \left(\frac{\partial S}{\partial u_{\alpha \beta}} \right)_T ~~~~~~~~~~~~~~~~~~~~ &&
\nonumber \\
   = \left(\frac{\partial E}{\partial u_{\alpha \beta}} \right)_S +
         T \left(\frac{\partial S}{\partial u_{\alpha \beta}} \right)_T =
         \left(\frac{\partial E}{\partial u_{\alpha \beta}} \right)_T~, && 
\label{dfdut}
\end{eqnarray}
where $T=(\partial E/\partial S)_{u_{\alpha \beta}}$ was used. 
(Keeping displacement gradients $u_{\alpha \beta}$ fixed ensures 
that the volume and shape of 
the crystal does not change.) Since
$F=E-TS$, Eq.\ (\ref{dfdut}) means that
\begin{equation}
         \left(\frac{\partial E}{\partial u_{\alpha \beta}} \right)_S =
         \left(\frac{\partial F}{\partial u_{\alpha \beta}} \right)_T~. 
\label{EeqF}
\end{equation}
Therefore 
\begin{eqnarray}
     \left(\frac{\partial^2 E}{\partial u_{\alpha \beta} \partial u_{\gamma \lambda}}\right)_S =
     \left[\frac{\partial}{\partial u_{\alpha \beta}} 
         \left(\frac{\partial F}{\partial u_{\gamma \lambda}} \right)_T \right]_S ~~~~~~~~~~~~~~~~~~~~~ &&
\nonumber \\
       =  \left(\frac{\partial^2 F}{\partial u_{\alpha \beta} \partial u_{\gamma \lambda}}\right)_T 
    + \left[\frac{\partial}{\partial T}  
         \left(\frac{\partial F}{\partial u_{\gamma \lambda}} \right)_T \right]_{u_{\alpha \beta}} 
         \left(\frac{\partial T}{\partial u_{\alpha \beta}} \right)_S~. &&   
\end{eqnarray}
Expressing derivative of $T$ from Eqs.\ (\ref{aux_derive}) and using (\ref{EeqF}) we obtain
\begin{eqnarray}
     \left(\frac{\partial^2 E}{\partial u_{\alpha \beta} \partial u_{\gamma \lambda}}\right)_S =
    \left(\frac{\partial^2 F}{\partial u_{\alpha \beta} \partial u_{\gamma \lambda}}\right)_T 
      ~~~~~~~~~~~~~ &&
\nonumber \\
     + ~T \left(\frac{\partial S}{\partial u_{\alpha \beta}}\right)_T 
        \left(\frac{\partial S}{\partial u_{\gamma \lambda}}\right)_T
    \left(\frac{\partial E}{\partial T }\right)^{-1}_{u_{\alpha \beta}}~. &&
\end{eqnarray}
Since $S=-(\partial F/\partial T)_{u_{\alpha \beta}}$, and 
$(\partial F/\partial u_{\alpha \beta})_T \propto \delta_{\alpha \beta}$
[i.e.\ zero for $\alpha \ne \beta$ and same for all $\alpha = \beta$;
cf.\ proof of Eq.\ (\ref{identity})],
we see that there is no difference between adiabatic and isothermal Huang 
coefficients $S_{1212}$ as well as $S_{1221}$. Also, we see that the difference
between adiabatic and isothermal coefficients $S_{1111}$ is the same as that 
for coefficients $S_{1122}$. This ensures that adiabatic and isothermal 
$\mu_{\rm eff}$ are the same.

\section{Numerical results}
\label{results}
In this section we present results of numerical calculations
of the elastic coefficients for the bcc lattice.
For such a lattice, one only needs to calculate four 
coefficients entering 
Eq.\ (\ref{mueff}), $S_{1111}$, $S_{1122}$, 
$S_{1221}$, and $S_{1212}$. All the other coefficients
with 2 pairs of identical indices are equal to these ones (e.g., 
$S_{1111}=S_{2222}=S_{3333}$,
$S_{1212}=S_{2121}=S_{1313}=\ldots\,$,
$S_{1221}=S_{2112}=S_{1331}=\ldots\,$, $S_{1122}=S_{2211}=S_{1133}=\ldots$),
while the rest of the coefficients are zero.
Alternatively, one can calculate $S_{1111}$, $S_{1122}$, $S_{1212}$, and
pressure $P$. In what follows we shall focus on static 
lattice and phonon contributions. As discussed 
in Sect.\ \ref{average}, there are no other contributions to the 
shear modulus.

The static lattice elastic coefficients are well-studied in 
the literature. 
Practical expressions 
for the coefficients of the expansion (\ref{U0expansion}) can 
be found in Appendix B. They are derived using standard Ewald technique
and given here for completeness.
The numerical results obtained using these 
expressions are given in Table \ref{fuchs}.
These values agree with the results of \citet{F36},
who calculated lattice energy expansion coefficients 
for two types of elastic deformations, 
$A = -P^{\rm st}+S^{\rm st}_{1111}-S^{\rm st}_{1122}$ and $2B = S^{\rm st}_{1212}$, for bcc and 
face-centered cubic
lattices \citep[see also][]{W67}.

\begin{table*}
\caption[]{Static lattice elastic coefficients (bcc) in units of $n Z^2 e^2 /(2 a_l)$,
where $a_l$ is the bcc lattice constant: $n a^3_l = 2$.}
\begin{tabular}{cccccc}
\hline \noalign{\smallskip}      
     $S^{\rm st}_{1111}$ & $S^{\rm st}_{1122}$ & $S^{\rm st}_{1212}$ & $-P^{\rm st}$  
   & $A$  & $\mu^{\rm st}_{\rm eff}$ \\ \noalign{\smallskip} 
\hline \noalign{\smallskip}
   ~$-1.4848079$~ & ~$-0.47067387$~ & ~$0.74240395$~ & 
                 ~$1.2130778$~ & ~$0.19894377$~ & ~$0.48523113$~ \\ \noalign{\smallskip} 
\hline
\end{tabular}
\label{fuchs}
\end{table*}

Computation of phonon coefficients (\ref{STabgl}) requires integration over the first 
Brillouin zone,
$\sum_{\bm{k}} = (2\pi)^{-3} V \int {\rm d} \bm{k}$.
Methods of such integration have been developed elsewhere
\citep{AG81,B00,BPY01}.
It is sufficient to integrate only
over the primitive cell of the Brillouin zone, described, for bcc lattice, 
by inequalities
$k_x \ge k_y \ge k_z \ge 0$ and $k_x+k_y \le 2 \pi/a_l$,
where $a_l$ is the bcc lattice constant: $n a_l^3 = 2$. 
The phonon frequencies are even functions of
$k_x,\,k_y,\,k_z$ and are invariant under an arbitrary permutation 
of $k_x,\,k_y,\,k_z$. 
In order to formulate symmetry properties of other quantities entering
(\ref{STabgl}) we denote an arbitrary permutation of $(x, y, z)$ as $(\rho, \sigma, \tau)$.
If one interchanges $k_\sigma \rightleftharpoons k_\tau$
then $\mathit{\Phi}^{jj}_{\rho \rho \rho \rho}$, 
$\mathit{\Phi}^{jj}_{\sigma \sigma \tau \tau}$ (equal to $\mathit{\Phi}^{jj}_{\tau \tau \sigma \sigma}$),
$\mathit{\Phi}^{jj}_{\sigma \tau \tau \sigma}$ (equal to $\mathit{\Phi}^{jj}_{\tau \sigma \sigma \tau}$)
remain the same, whereas 
$\mathit{\Phi}^{jj}_{\sigma \sigma \sigma \sigma} \rightleftharpoons \mathit{\Phi}^{jj}_{\tau \tau \tau \tau}$, 
$\mathit{\Phi}^{jj}_{\rho \rho \sigma \sigma} \rightleftharpoons \mathit{\Phi}^{jj}_{\rho \rho \tau \tau}$, 
$\mathit{\Phi}^{jj}_{\rho \sigma \sigma \rho} \rightleftharpoons \mathit{\Phi}^{jj}_{\rho \tau \tau \rho}$, 
$\mathit{\Phi}^{jj}_{\rho \sigma \rho \sigma} \rightleftharpoons \mathit{\Phi}^{jj}_{\rho \tau \rho \tau}$, 
$\mathit{\Phi}^{jj}_{\sigma \rho \sigma \rho } \rightleftharpoons \mathit{\Phi}^{jj}_{\tau \rho \tau \rho}$, 
and
$\mathit{\Phi}^{jj}_{\sigma \tau \sigma \tau} \rightleftharpoons \mathit{\Phi}^{jj}_{\tau \sigma \tau \sigma}$.
The same relationships govern symmetry properties of products 
$\mathit{\Phi}^{jj'}_{\alpha \beta} \mathit{\Phi}^{jj'}_{\gamma \lambda}$,
viewed as 4-index ($\alpha \beta \gamma \lambda$) quantities. Additionally, 
$\mathit{\Phi}^{jj}_{\sigma \sigma \tau \tau} = \mathit{\Phi}^{jj}_{\sigma \tau \tau \sigma}$,
and
$\mathit{\Phi}^{jj}_{\sigma \tau} = \mathit{\Phi}^{jj}_{\tau \sigma}$,
so that 
$\mathit{\Phi}^{jj}_{\sigma \tau} \mathit{\Phi}^{jj}_{\sigma \tau} = 
\mathit{\Phi}^{jj}_{\tau \sigma} \mathit{\Phi}^{jj}_{\tau \sigma} =
\mathit{\Phi}^{jj}_{\sigma \tau} \mathit{\Phi}^{jj}_{\tau \sigma}$.
Both 
$\mathit{\Phi}^{jj}_{\alpha \beta \gamma \lambda}$
and products 
$\mathit{\Phi}^{jj'}_{\alpha \beta} \mathit{\Phi}^{jj'}_{\gamma \lambda}$
are even functions of $k_x,\,k_y,\,k_z$.

Let us give the recipe to calculate an arbitrary $S^{\rm ph}$ coefficient from Eq.\ (\ref{STabgl}).
In order to obtain, for instance, $S^{\rm ph}_{1111}$ one has to integrate over the
primitive cell of the Brillouin zone the integrand in Eq.\ (\ref{STabgl}) 
with ${\alpha \beta \gamma \lambda} = {xxxx},\,{yyyy},$ and ${zzzz}$. Then add
the three results together and multiply by 16 to account for the remaining two
permutations for positive $k_x,\,k_y,\,k_z$ and for 8 possible combinations
of signs of $k_x,\,k_y,\,k_z$. The same recipe is applied in the case of
$S^{\rm ph}_{1122}$ and $S^{\rm ph}_{1221}$. 

The situation with $S^{\rm ph}_{1212}$ 
is
more complex because,
for instance, $\mathit{\Phi}^{jj}_{xyxy} \ne \mathit{\Phi}^{jj}_{yxyx}$. 
This 
requires 
integration over the primitive cell of the 
integrand of Eq.\ (\ref{STabgl}) with 
${\alpha \beta \gamma \lambda} = {xyxy},\,{yxyx},\,{xzxz},\,{zxzx},\,{yzyz},$ 
and ${zyzy}$, addition of all 6 of them together and multiplication by 8.

\begin{figure}
\begin{center}
\leavevmode
\includegraphics[height=84mm,bb=12 3 344 349,clip]{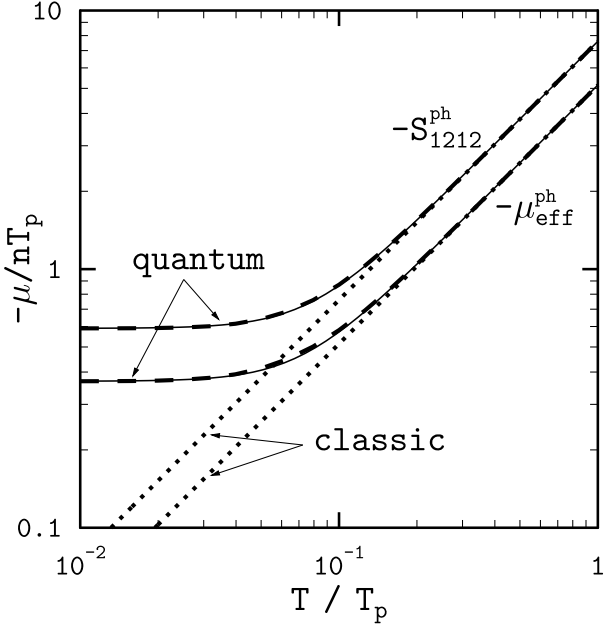}
\end{center}
\vspace{-0.4cm} \caption[ ]{Phonon elastic coefficient $-S^{\rm ph}_{1212}$ (upper dashed curve)
and effective shear modulus $-\mu^{\rm ph}_{\rm eff}$ (lower dashed curve) 
in units of $n T_{\rm p}$ vs.\ $T/T_{\rm p}$. Solid curves and dots  
show analytic fits [Eqs.\ (\ref{fit}) and (\ref{fitS})] and classic numerical results, respectively.} 
\label{mu_ph}
\end{figure}
%

We shall now describe
the results of numerical calculations. 
In Fig.\ \ref{mu_ph} we show $-S^{\rm ph}_{1212}$ (upper dashed curve) and 
$-\mu^{\rm ph}_{\rm eff}$ (lower dashed curve)
in units of $n T_{\rm p}$ as functions of $T/T_{\rm p}$, where
$T_{\rm p} = \hbar \omega_{\rm p}$ is the ion plasma temperature.
Quantities $S^{\rm ph}_{1212}$ and $\mu^{\rm ph}_{\rm eff}$
are negative. Thus they reduce the respective static lattice values 
(Table \ref{fuchs}) and weaken lattice resistance to the shear strain.
Dots show the same quantities with quantum effects 
explicitly excluded. In this case the elastic coefficients are always proportional to $T$.
These results are obtained by setting 
$\bar{n}_{\bm{k}j} = T/\hbar \omega_{\bm{k}j}$ and retaining only the
highest order terms in $T$ in Eq.\ (\ref{STabgl}). 
At higher temperatures
the classic curves merge with the exact results, whereas
at lower temperatures the quantum effects dominate. 
The exact curves do not decrease beyond certain values corresponding
to ion zero-point motion.  
In this temperature regime the perturbation theory has clear advantage
over Monte Carlo or molecular dynamics methods, in which the ion motion 
is treated classically. 

We were able to fit the phonon shear coefficients as
\begin{eqnarray}
       \mu^{\rm ph}_{\rm eff} &=& - n T_{\rm p} \, \left[ 0.3686^3 + 
            136.6 \left( \frac{T}{T_{\rm p}} \right)^3 \right]^{1/3}~,
\label{fit} \\
       S^{\rm ph}_{1212} &=& - n T_{\rm p} \, \left[ 0.5903^3 + 
            439 \left( \frac{T}{T_{\rm p}} \right)^3 \right]^{1/3}~.
\label{fitS}
\end{eqnarray}
These curves are shown in Fig.\ \ref{mu_ph} as thin solid lines 
(merging with the dashed ones). 
They reproduce exactly the classic and zero-point limits. 
The maximum error of the $\mu^{\rm ph}_{\rm eff}$ fit 
is 2.2\% at $T/T_{\rm p} \approx 0.06$.
The maximum error of the $S^{\rm ph}_{1212}$ fit 
is 1.0\% at $T/T_{\rm p} \approx 0.08$.

\begin{figure}
\begin{center}
\leavevmode
\includegraphics[height=84mm,bb=11 3 351 345,clip]{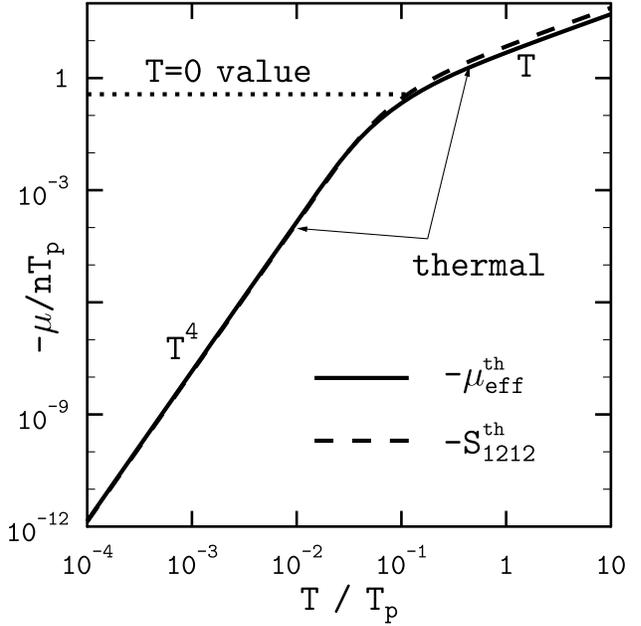}
\end{center}
\vspace{-0.4cm} \caption[ ]{Thermal contributions
$-\mu^{\rm th}_{\rm eff}$ (solid line) and $-S^{\rm th}_{1212}$ (dashed line) and
zero-point contribution to $-\mu^{\rm ph}_{\rm eff}$ (dots) in units of 
$n T_{\rm p}$ vs.\ $T/T_{\rm p}$.} 
\label{mu_th}
\end{figure}
%

If one subtracts the $T \to 0$ limit (i.e.\ the zero-point contribution) 
from the phonon elastic coefficients, 
the remaining part will correspond to thermal ion motion 
(we denote this part by a superscript `th').
In analogy with the derivation of the Debye $T^3$-law for specific 
heat, one can show that at low $T$ this thermal contribution to the 
elastic coefficients behaves as $T^4$ \citep[e.g.,][]{BH54}. In the classic regime of high
$T$, both $S^{\rm th}_{1212}$ and $\mu^{\rm th}_{\rm eff}$ are, naturally, proportional to
$T$ (cf.\ Fig.\ \ref{mu_ph}). Our numerical calculations of $-\mu^{\rm th}_{\rm eff}$ 
and $-S^{\rm th}_{1212}$ reproduce both asymptotes and are shown in Fig.\ \ref{mu_th}
by solid and dashed lines, respectively. Dots 
show the zero-point
contribution to $\mu^{\rm ph}_{\rm eff}$. Since it is the 
sum of the zero-point and thermal contributions that make up the total phonon elastic coefficient, 
the $T^4$ part of the thermal contribution is practically always negligible. 
We summarize parameters of various asymptotes in Table \ref{asympt-param}.

\begin{table}
\caption[]{Asymptote parameters $f$, $g$, and $h$ in units of $nT_{\rm p}$, where $f(T/T_{\rm p})^4$ is
the thermal quantum asymptote, $gT/T_{\rm p}$ is the classic asymptote, 
and $h$ is the $T=0$ (zero-point) value of $-\mu^{\rm ph}_{\rm eff}$ and $-S^{\rm ph}_{1212}$.}
\begin{eqnarray}
\begin{tabular}{c|ccc}
    & $f$ &  $g$ &  $h$  \\ \hline \\[-0.2cm]
  $-\mu^{\rm ph}_{\rm eff}$ & ~$1.43 \times 10^4$~ & ~$5.14$~ & ~$0.369$ \\[0.1cm]
  $-S^{\rm ph}_{1212}$ & ~$1.35 \times 10^4$~ & ~$7.60$~ & ~$0.590$ \\
\end{tabular}
\nonumber
\end{eqnarray}
\label{asympt-param}
\end{table}

\begin{figure}
\begin{center}
\leavevmode
\includegraphics[height=82mm,bb=3 14 347 347,clip]{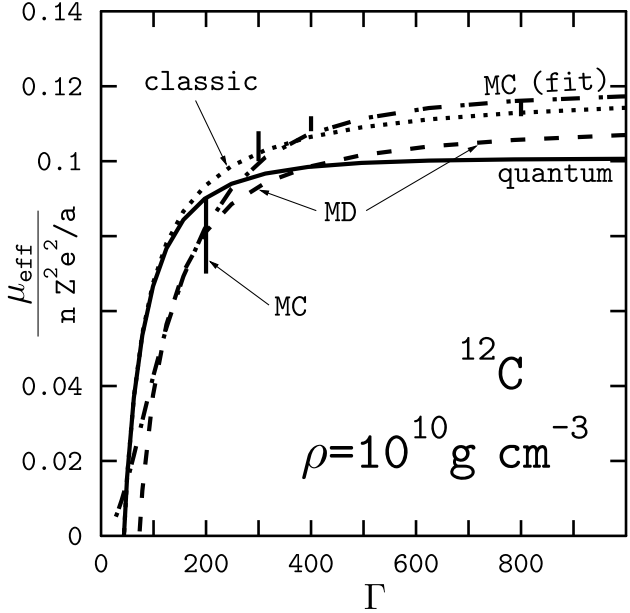}
\end{center}
\vspace{-0.4cm} \caption[ ]{Total effective shear modulus 
in units of $n Z^2 e^2 /a$ vs.\ $\Gamma$ for fully ionized
$^{12}$C at density $10^{10}$ g cm$^{-3}$. Solid line and dots
represent results of our quantum and classic calculations, respectively,
bars show Monte Carlo (MC) data, dash-dotted curve is the fit to them, and
dashed curve represents molecular dynamics (MD) results.} 
\label{mu_tot}
\end{figure}

Finally, let us compare our results with those of other authors.
In Fig.\ \ref{mu_tot} we show various approximations to the total
effective shear modulus in units of $n Z^2 e^2 /a$ versus $\Gamma = Z^2 e^2 /(aT)$
for fully ionized $^{12}$C at density $10^{10}$ g cm$^{-3}$. 
In this case $a = (4 \pi n/3)^{-1/3}$ is the ion-sphere radius and $\Gamma$
is the standard Coulomb coupling parameter. Bars represent original 
Monte Carlo calculations of \citet{OI90}, while dash-dotted curve
shows the fit to these data from \citet{SVHO91}. The solid line is 
$\mu^{\rm st}_{\rm eff} + \mu^{\rm ph}_{\rm eff}$, where 
$\mu^{\rm st}_{\rm eff}$ is from Table \ref{fuchs} and $\mu^{\rm ph}_{\rm eff}$ is given 
by Eq.\ (\ref{fit}). Dots show the sum of $\mu^{\rm st}_{\rm eff}$
and the classic asymptote of $\mu^{\rm ph}_{\rm eff}$ 
(dots in Fig.\ \ref{mu_ph}). This curve thus represents results of a purely 
classic calculation and may be directly compared with the Monte Carlo study.
The dashed line shows results of molecular dynamics simulations reported 
by \citet{HH08}.

%

First of all, we note that our classic calculation (dots) matches the
bars of \citet{OI90} with the exception of the one at $\Gamma=200$, where there is 
a discrepancy of about 3\% between the dotted curve and the upper tip of the error bar.
One possible reason for this discrepancy is associated with anharmonic corrections 
to the free energy, not taken into account in our perturbative calculation. At 
$\Gamma=200$ the ratio of anharmonic and harmonic energies of the classic crystal is 
$(A_1 N T/\Gamma)/(3 N T) \approx 1.8\%$, where $A_1 \approx 10.84$ \citep{D90}.
On the other hand, it is not clear from \citet{OI90} what is the confidence 
level of their error bars.

The difference between the quantum calculations (solid curve) and the classic ones depends on
temperature and reflects the results shown in Fig.\ \ref{mu_ph}. 
The deviation of the quantum curve is mostly due to zero-point 
ion vibrations and is the strongest at lower temperatures (higher $\Gamma$'s). The relative magnitude of the 
deviation is proportional to $a \hbar \omega_p/(Z^2 e^2) \approx 0.05 \rho_{10}^{1/6}/(Z_6 A_{12}^{2/3})$,
where $Z_6 = Z/6$, $A_{12} = A/12$, and $\rho_{10}$ is the mass density in units of
$10^{10}$ g cm$^{-3}$. For carbon at $\rho = 10^{10}$ g cm$^{-3}$ and low $T$'s, the classic effective
shear modulus exceeds the more accurate quantum result by up to 18\%. 

As expected, at higher temperatures the quantum curve merges 
with the classic one. It is well known that for a Coulomb system the liquid state is energetically preferable
to the crystal at $\Gamma \la 175$. If there is a crystal at these $\Gamma$'s, 
it is in the metastable overheated state.
We have extended our calculations into this temperature range for the sake of a
qualitative discussion. Obviously, the further
we go into the overheated crystal regime, the less accurate our results become due to the growing
importance of anharmonic effects. In accordance
with the general picture of Fig.\ \ref{mu_ph}, where the phonon contribution into the
effective shear modulus is negative and grows by absolute value with temperature, the total 
effective shear modulus
drops and, eventually, reaches zero at $\Gamma \sim 45$. This point should be regarded as a lower
bound for the disappearance of the shear modulus in a Coulomb system. 
The actual crossing point is expected to 
occur at higher $\Gamma$'s and will be sensitive to the anharmonic effects. 

Molecular dynamics calculations of the effective shear modulus in a Coulomb system with
compressible electron background, producing screening, were performed by
\citet{HH08}. These results are shown in Fig.\ \ref{mu_tot} by the dashed line. 
The electron screening was described by the Thomas-Fermi (TF) model. By its nature the 
molecular dynamics approach does not take into account quantum effects.
We observe a decrease of the shear modulus (with respect to the classic dotted curve) 
reflecting the fact 
that the crystal with screened Coulomb potential is less tightly bound than the pure Coulomb crystal. 

It is well known that to lowest order in electromagnetic coupling,
the screening by relativistic degenerate electrons is described by the (RPA) dielectric 
function of \citet{J62}.
The resulting effective ion-ion potential is more complex than the simple Yukawa potential arising in
the TF model. Electron screening affects both properties of the static lattice and
the phonon modes. As shown by \citet{B02}, the TF model overestimates the correction to the lattice
electrostatic energy as compared to the more accurate RPA model. By contrast, the properties
of the phonon modes obtained in the TF and RPA models are very similar. 
Since the main contribution to the effective shear modulus comes from the electrostatic energy,
we expect, that calculations of \citet{HH08} in the TF model overestimate the importance of 
the electron background compressibility. It would be interesting (and is not difficult) to calculate the 
electrostatic contribution to the effective shear modulus using the full RPA model.

\section{Conclusions}
We have calculated elastic shear coefficient $S_{1212}$ and effective direction averaged
shear modulus $\mu_{\rm eff}$ for neutron star crust matter. Both coefficients
are sums of a static lattice term and a term originating
from the motion of ions about their lattice nodes. While the static lattice 
contributions (Table \ref{fuchs}) were well known previously, only 
numerical simulations existed for the ion motion terms.
Using thermodynamic perturbation theory, we have expressed the ion motion terms
via integrals over the first Brillouin zone of quantities given by rapidly convergent 
lattice sums, Eq.\ (\ref{STabgl}). The integrals were then evaluated numerically and 
the results were fitted by simple analytic formulae, 
Eqs.\ (\ref{fit}) and (\ref{fitS}).

The main advantage of the numerical methods (Monte Carlo and molecular dynamics)
is their ability to include anharmonic effects to all orders. The main advantage of the 
perturbation theory is its ability to include quantum effects (within the framework 
of the harmonic lattice model), greater transparency of the results and 
much lesser computer time requirements. The anharmonic effects can also be taken 
into account in the perturbative approach, but that would make all the equations much 
more cumbersome even if only the lowest order anharmonic term is retained.
Summarizing, we can say that numerical simulations and perturbation theory 
ideally complement each other as well as serve for mutual verification.

If quantum effects are included, one finds that the ion motion contribution can be decomposed
into two parts. One of them corresponds to zero-point ion motion and is independent of $T$.
The other corresponds to thermal ion motion and is $\propto T^4$ at low temperatures
and $\propto T$ at high temperatures. The high $T$ asymptote is what one obtains, 
if the calculation is purely classic. At low temperatures the thermal term 
is negligible compared to the zero-point term, and thus the $T^4$ asymptote 
seems rather unimportant. Our fitting formulae (\ref{fit}) and (\ref{fitS})
reproduce exactly the high $T$ and zero-point limits.

If quantum effects are excluded, our results agree well (cf.\ dots and bars in Fig.\ \ref{mu_tot})
with Monte Carlo simulations of \citet{OI90}. The only discrepancy of about 3\% occurs at 
$\Gamma=200$, where anharmonic effects are at their strongest (and also the error bars of 
the Monte Carlo simulation are at their largest). If quantum effects are included,
then the main difference with the Monte Carlo results is due to the zero-point contribution
to the ion motion term. Compared to the total shear modulus, that also includes the 
static lattice part, this contribution is important for lighter elements at higher densities,
where the ion plasma temperature is not entirely negligible with respect 
to the typical Coulomb ion interaction energy. 

We have demonstrated that neither $S_{1212}$ nor $\mu_{\rm eff}$ have any contributions from
subsystems whose partial free energies are functions of particle number density only; that 
both coefficients are the same whether they are evaluated at constant temperature or entropy;
we have also proven an identity linking phonon pressure with a coefficient of 
dynamic matrix expansion in powers of displacement gradients, Eq.\ (\ref{identity}).

The results, reported in this paper, also apply to crystallized matter in white dwarf cores.

\section*{Acknowledgments}
The author is grateful to D. G. Yakovlev for suggesting the topic 
of this research and numerous fruitful discussions.
The work was supported by Ministry of Education and Science of Russian Federation 
(contract No. 11.G34.31.0001 with SPbSPU and leading scientist G. G. Pavlov), 
by RFBR (grant 11-02-00253-a) and by Rosnauka (grant NSh 3769.2010.2).

\appendix

\section{}
Both $D_{\alpha \beta}^{\mu \nu}(\bm{k})$ and 
$D_{\alpha \beta \gamma \lambda}^{\mu \nu}(\bm{k})$
are sums of two series, over direct and reciprocal
lattice vectors (denoted ${\bm R}$ and ${\bm G}$ respectively):  
$D_{\alpha \beta}^{\mu \nu}=D_{\alpha \beta}^{\mu \nu [R]}+D_{\alpha \beta}^{\mu \nu [G]}$,
$D_{\alpha \beta \gamma \lambda}^{\mu \nu}=D_{\alpha \beta \gamma \lambda}^{\mu \nu [R]}+
D_{\alpha \beta \gamma \lambda}^{\mu \nu [G]}$.
\begin{eqnarray}
       D_{\alpha \beta}^{\mu \nu [R]} (\bm{k}) &=& \frac{Z^2 e^2}{M}
             \sideset{}{'}\sum_{\bm{R}} [1 - \exp{({\rm i} \bm{k} \cdot \bm{R})}] 
\nonumber \\
        &\times&
         \left\{\frac{4}{R^5} 
            \left( \delta_{\alpha \mu} R_\beta R_\nu + \delta_{\alpha \nu} 
            R_\beta R_\mu + 
             \delta_{\mu \nu} R_\alpha R_\beta \right) \right.
\nonumber \\
        &\times& \left[ 
            \left( A^3 R^3 + \frac{3}{2} AR \right) 
          \frac{e^{-A^2 R^2}}{\sqrt{\pi}} + 
           \frac{3}{4} \, {\rm erfc}(AR) \right] 
\nonumber \\
        &-& \frac{8 R_\alpha R_\beta R_\mu R_\nu}{R^7} 
\nonumber \\
          &\times& 
        \left[  \left( A^5 R^5 +\frac{5}{2} A^3 R^3 
         +  \frac{15}{4} AR \right)
           \frac{e^{-A^2 R^2}}{\sqrt{\pi}} \right.
\nonumber \\
   &+& \left. \left. \frac{15}{8} \, {\rm erfc}(AR) \right] \right\}~,
\label{DR(ab)}
\end{eqnarray}
\begin{eqnarray}
       D_{\alpha \beta}^{\mu \nu [G]} (\bm{k}) &=& \frac{4 \pi n Z^2 e^2}{M} 
        \int \frac{{\rm d}\bm{q}}{q^2}  
          \left[ \sideset{}{'}\sum_{\bm{G}} \delta (\bm{q} - \bm{G}) \right.
\nonumber \\
          &-& \left. \sum_{\bm{G}} \delta (\bm{q} - \bm{k} - \bm{G}) \right] 
          e^{-q^2/(4A^2)}
\nonumber \\
       &\times&  \left[ \delta_{\alpha \beta} q_\mu q_\nu +
     \delta_{\mu \beta} q_\alpha q_\nu + \delta_{\nu \beta} q_\alpha q_\mu \right.
\nonumber \\
  &-& \left.
       \frac{2}{q^2} \left( 1 + \frac{q^2}{4A^2} \right) 
           q_\alpha q_\beta q_\mu q_\nu  \right]~,
\label{DG(ab)}
\end{eqnarray}
\begin{eqnarray}
       D_{\alpha \beta \gamma \lambda}^{\mu \nu [R]} (\bm{k}) &=& 
            \frac{Z^2 e^2}{M}
             \sideset{}{'}\sum_{\bm{R}} [1 - \exp{({\rm i} \bm{k} \cdot \bm{R})}] 
\nonumber \\
           &\times&
           \left\{\frac{4 R_\beta R_\lambda}{R^5} 
            \left( \delta_{\alpha \gamma} \delta_{\mu \nu} + 
               \delta_{\alpha \mu} \delta_{\gamma \nu} +
             \delta_{\alpha \nu} \delta_{\gamma \mu} \right) \right.
\nonumber \\
           &\times&  \left[ 
            \left( A^3 R^3 + \frac{3}{2} AR \right) 
             \frac{e^{-A^2 R^2}}{\sqrt{\pi}} + 
           \frac{3}{4} \, {\rm erfc}(AR) \right]
\nonumber \\
           &-& \frac{8 R_\beta R_\lambda}{R^7}
               \left( \delta_{\alpha \gamma} R_\mu R_\nu+ 
                      \delta_{\alpha \mu} R_\gamma R_\nu+
                      \delta_{\alpha \nu} R_\gamma R_\mu \right.
\nonumber \\
                 &+& \left. \delta_{\gamma \mu} R_\alpha R_\nu+
                \delta_{\gamma \nu} R_\alpha R_\mu+
               \delta_{\mu \nu} R_\alpha R_\gamma \right)
\nonumber \\
        &\times& \left[  \left( A^5 R^5 +\frac{5}{2} A^3 R^3 + 
          \frac{15}{4} AR \right)
           \frac{e^{-A^2 R^2}}{\sqrt{\pi}} \right.
\nonumber \\
       &+& \left.  \frac{15}{8} \, 
              {\rm erfc}(AR) \right]
\nonumber \\
         &+& \frac{16 R_\alpha R_\beta R_\gamma R_\lambda R_\mu R_\nu}{R^9}
\nonumber \\
         &\times&   \left[  \left( A^7 R^7 + \frac{7}{2} A^5 R^5 
          + \frac{35}{4} A^3 R^3  + \frac{105}{8} AR \right) \right.
\nonumber \\
          &\times&
           \frac{e^{-A^2 R^2}}{\sqrt{\pi}} \left. \left.
                 +  \frac{105}{16} \, {\rm erfc}(AR) \right] \right\}~,
\label{DR(ab)(gl)} 
\end{eqnarray}
\begin{eqnarray}
       D_{\alpha \beta \gamma \lambda}^{\mu \nu [G]} (\bm{k}) &=&  
             - \frac{4 \pi n Z^2 e^2}{M} 
        \int \frac{{\rm d}\bm{q}}{q^2}  
          \left[ \sideset{}{'}\sum_{\bm{G}} \delta (\bm{q} - \bm{G}) \right.
\nonumber \\
          &-& \left. \sum_{\bm{G}} \delta (\bm{q} - \bm{k} - \bm{G}) \right] 
           e^{-q^2/(4A^2)} 
\nonumber \\
           &\times&
       \left[ \delta^{(2)} q^{(2)}  - \frac{2}{q^2} 
                \left( 1 + \frac{q^2}{4A^2} \right)
              \delta q^{(4)} \right.
\nonumber \\
               &+& \left. \frac{8}{q^4} 
                \left( 1 + \frac{q^2}{4A^2} + 
             \frac{q^4}{32A^4} \right) q^{(6)} \right]~,
\label{DG(ab)(gl)}
\end{eqnarray}
where
\begin{eqnarray}
           \delta^{(2)} q^{(2)} &=& 
             \delta_{\alpha \lambda} \left( \delta_{\gamma \beta} 
                q_\mu q_\nu +
               \delta_{\mu \beta} q_\gamma q_\nu +
                \delta_{\nu \beta} q_\gamma q_\mu \right) 
\nonumber \\
&+&
             \delta_{\gamma \lambda} \left( \delta_{\alpha \beta} 
             q_\mu q_\nu +
               \delta_{\mu \beta} q_\alpha q_\nu +
                \delta_{\nu \beta} q_\alpha q_\mu \right)
\nonumber \\
          &+&   \delta_{\mu \lambda} \left( \delta_{\alpha \beta} 
             q_\gamma q_\nu +
               \delta_{\gamma \beta} q_\alpha q_\nu +
                \delta_{\nu \beta} q_\alpha q_\gamma \right)
\nonumber \\
              &+&    \delta_{\nu \lambda} \left( \delta_{\alpha \beta} 
               q_\gamma q_\mu +
               \delta_{\gamma \beta} q_\alpha q_\mu +
                \delta_{\mu \beta} q_\alpha q_\gamma \right)~,
\label{d2q2} \\
          \delta q^{(4)} &=& \delta_{\alpha \lambda} 
               q_\beta q_\gamma q_\mu q_\nu +
                                     \delta_{\beta \lambda} 
               q_\alpha q_\gamma q_\mu q_\nu +
                                     \delta_{\gamma \lambda} 
               q_\alpha q_\beta q_\mu q_\nu 
\nonumber \\
                  &+&
                                     \delta_{\mu \lambda} 
               q_\alpha q_\beta q_\gamma q_\nu +
                                     \delta_{\nu \lambda} 
               q_\alpha q_\beta q_\gamma q_\mu
\nonumber \\
             &+& \delta_{\alpha \beta}  q_\gamma q_\mu q_\nu q_\lambda 
              +   \delta_{\gamma \beta}  q_\alpha q_\mu q_\nu q_\lambda 
\nonumber \\
             &+&
                 \delta_{\mu \beta}  q_\alpha q_\gamma q_\nu q_\lambda 
                 +
                 \delta_{\nu \beta}  q_\alpha q_\gamma q_\mu q_\lambda~,
\label{dq4} \\ 
          q^{(6)} &=& q_\alpha q_\beta q_\gamma q_\lambda q_\mu q_\nu~.
\label{q6}
\end{eqnarray}
In the above equations, as well as in Appendix B, erfc is the complementary error
function, and $A$ is an arbitrary parameter with units of inverse length
chosen so that sums over direct and reciprocal lattice vectors are 
equally rapidly convergent. Integrals over ${\rm d}\bm{q}$ in Eqs.\ (\ref{DG(ab)})
and (\ref{DG(ab)(gl)}) are kept for brevity of the formulae. They can be
easily evaluated with the aid of the $\delta$-functions.

\section{}
Similarly to Appendix A, $S^{\rm st}_{\alpha \beta}=S_{\alpha \beta}^{[R]}+S_{\alpha \beta}^{[G]}$
and $S^{\rm st}_{\alpha \beta \gamma \lambda}=S_{\alpha \beta \gamma \lambda}^{[R]}+
S_{\alpha \beta \gamma \lambda}^{[G]}$, where
\begin{eqnarray}
              S_{\alpha \beta}^{[R]} &=& - n Z^2 e^2 \sideset{}{'}\sum_{\bm{R}} 
             \frac{R_\alpha R_\beta}{R^3}
           \left[ \frac{AR}{\sqrt{\pi}} e^{-A^2 R^2} + \frac{1}{2} 
            {\rm erfc}(AR) \right]
\nonumber \\
          &+& \frac{n^2 \pi Z^2 e^2}{2 A^2} \delta_{\alpha \beta}~,
\label{SR(ab)} 
\end{eqnarray}
\begin{eqnarray}
             S_{\alpha \beta}^{[G]} &=& - 2 \pi n^2 Z^2 e^2 \sideset{}{'}\sum_{\bm{G}}  
               \frac{1}{G^2} 
\nonumber \\ 
           &\times& \left[ \delta_{\alpha \beta} - 
              \frac{2}{G^2} \left( 1 + \frac{G^2}{4 A^2}\right) 
            G_\alpha G_\beta \right]
               e^{-G^2/(4A^2)}~,
\label{SG(ab)}
\end{eqnarray}
\begin{eqnarray}
            S_{\alpha \beta \gamma \lambda}^{[R]} &=& 
            n Z^2 e^2 \sideset{}{'}\sum_{\bm{R}} \left\{ 
             \frac{2 R_\alpha R_\beta R_\gamma R_\lambda}{R^5} \right.
\nonumber \\
            &\times& \left[ 
            \left( A^3 R^3 + \frac{3}{2} AR \right) 
             \frac{e^{-A^2 R^2}}{\sqrt{\pi}} + 
           \frac{3}{4} \, {\rm erfc}(AR) \right]
\nonumber \\
            &-& \left. \delta_{\alpha \gamma} \frac{R_\beta R_\lambda}{R^3} 
           \left[ \frac{AR}{\sqrt{\pi}} e^{-A^2 R^2} + 
           \frac{1}{2} {\rm erfc}(AR) \right] \right\}
\nonumber \\
           &-& \frac{n^2 \pi Z^2 e^2}{2 A^2} \left( \delta_{\alpha \beta} 
          \delta_{\gamma \lambda} + 
                 \delta_{\alpha \lambda} \delta_{\beta \gamma} \right)~,
\label{SR(ab)(gl)} 
\end{eqnarray}
\begin{eqnarray}
             S_{\alpha \beta \gamma \lambda}^{[G]} &=& 2 \pi n^2 Z^2 e^2
             \sideset{}{'}\sum_{\bm{G}}  
               \frac{1}{G^2} \left[ \left( \delta_{\alpha \beta} 
                \delta_{\gamma \lambda} + 
                 \delta_{\alpha \lambda} \delta_{\beta \gamma} \right) \right.
\nonumber \\
    &-& \frac{2}{G^2} \left( 1 + \frac{G^2}{4 A^2}\right) 
                \left( \delta_{\alpha \beta} G_\gamma G_\lambda
                       + \delta_{\beta \gamma} G_\alpha G_\lambda \right.
\nonumber \\
                 &+& \left. \delta_{\beta \lambda} G_\alpha G_\gamma
                       + \delta_{\alpha \lambda} G_\beta G_\gamma
                       + \delta_{\gamma \lambda} G_\alpha G_\beta \right) 
\nonumber \\
              &+& \left. \frac{8}{G^4} \left(1 +  \frac{G^2}{4A^2} + 
             \frac{G^4}{32A^4} \right) 
            G_\alpha G_\beta G_\gamma G_\lambda \right] 
\nonumber \\ 
           &\times& e^{-G^2/(4A^2)}~.
\label{SG(ab)(gl)}
\end{eqnarray}

\label{lastpage}

\end{document}